\documentclass[british,letterpaper,aps,pre,twocolumn,showcaps,superscriptaddress,amsmath,amssymb]{revtex4}
\usepackage{float}
\usepackage{graphicx}
\usepackage{amssymb}
\usepackage{color}
\makeatletter

\usepackage{babel}
\makeatother

\begin{document}

\title{The nature of the glass and gel transitions in sticky spheres}

\author{C. Patrick Royall}

\affiliation{HH Wills Physics Laboratory, Tyndall Avenue, Bristol, BS8 1TL, UK.}
\affiliation{School of Chemistry, University of Bristol, Cantock Close, Bristol, BS8 1TS, UK.}
\affiliation{Centre for Nanoscience and Quantum Information, Tyndall Avenue, Bristol, BS8 1FD, UK.}

\author{Stephen R. Williams}

\affiliation{Research School of Chemistry, Australian National University, Canberra, ACT 0200, Australia.}

\author{Hajime Tanaka}

\affiliation{Institute of Industrial Science, University of Tokyo, 4-6-1 Komaba, Meguro-ku, Tokyo, 153-8505, Japan.}

\date{\today}

\begin{abstract}
Glasses \cite{berthier2011} and gels \cite{zaccarelli2007} are the two dynamically arrested, disordered states of matter. Despite their importance, their similarities and differences remain elusive, especially at high density. We identify dynamical and structural signatures which distinguish the gel and glass transitions in a colloidal model system of hard and ``sticky'' spheres. Gelation is induced by crossing the gas-liquid phase-separation line \cite{verhaegh1997,tanaka1999colloid,lu2008} and the resulting rapid densification of the colloid-rich phase leads to a sharp change in dynamics and local structure.  Thus, we find that gelation is first-order-like and can occur at much higher densities than previously thought: far from being low-density networks, gels have a clear ``thermodynamic'' definition which nevertheless leads to a non-equilibrium state with a distinct local structure characteristic of a rapidly quenched glass. In contrast, approaching the glass transition, the dynamics slow continuously accompanied by the emergence of local five-fold symmetric structure. Our findings provide a general thermodynamic, kinetic, and structural basis upon which to distinguish gelation from vitrification.
\end{abstract}
\maketitle

\section{Introduction}
\label{sectionIntroduction}

Dynamical arrest remains one of the principle unsolved challenges in condensed matter \cite{berthier2011}. In everyday soft materials, arrest takes two forms: vitrification, the process by which a fluid becomes a glass, and gelation, a transition to the gel state \cite{poon2002,ramos2005sdg,zaccarelli2007}. A glass is known to have spatially homogeneous density on lengthscales larger than a few particle diameters. On the other hand, a typical gel has a sparse percolated network structure at low overall colloid concentration. However, with an increase in concentration the network structure becomes thicker and the system becomes more homogeneous in density, which makes the distinction between gels and glasses increasingly obscure. The comparison between these non-ergodic states has so far been dominated by considerations of glasses and gels with very different structures. For elucidating the fundamental difference between the glass transition and gelation, however, the critical comparison should be made for a gel and a glass with a similar density. Since a glass is formed only at a high density, the comparison should be made at high density. At such high density, however, both form similar disordered non-ergodic structures, which are rather homogeneous in density beyond the particle scale. This makes the comparison very challenging. While it is possible to make a dynamical distinction between glasses and gels using rheology \cite{pham2006,koumakis2011} or simulation \cite{zaccarelli2009,puertas2002}, these measures lead to a crossover in behaviour, rather than a sharp transition. This is due to the emphasis on the interplay of caging (glass) and bonding (gel) behaviour. Since the onset of vitrification is continuous \cite{berthier2011}, the interplay between these effects is also continuous.

Here we take a different approach. By considering thermodynamic properties, we identify a clear transition between gelation and vitrification which reveals fundamental differences between the two types of ergodic-nonergodic transition. These lead us to address questions such as what physical factors allow us to distinguish these two non-ergodic states and whether the transition between them is continuous or discontinuous. To do so we focus on a system of colloidal ``sticky spheres''. In the absence of attraction, colloidal ``hard'' spheres undergo a glass transition upon increasing volume fraction \cite{brambilla2009,poon2002,ramos2005sdg}. The addition of polymer leads to an effective attraction between the colloids due to polymer depletion. The range and strength of this interaction are set by the polymer size and concentration respectively. The depth of the attractive well in this polymer-mediated colloid-colloid interaction sets an effective temperature. Upon increasing the polymer concentration the effective temperature is reduced and systems with short-ranged (sticky) attractions undergo gelation at moderate colloid volume fractions \cite{poon2002,zaccarelli2007}. This gelation is believed to be related to arrested spinodal decomposition to a (colloidal) liquid and gas \cite{tanaka1999colloid,lu2008}. At high volume fraction, the hard sphere glass melts upon the introduction of attractions and the resulting (ergodic) fluid subsequently undergoes re-entrant arrest upon increasing the attraction strength \cite{pham2002,zaccarelli2002,poon2002}. 

The transition between a supercooled liquid and a glass 
occurs when structural relaxation can no longer be observed on the experimental timescale. At lower temperatures (or higher densities) lack of equilibration makes it hard to determine the existence or otherwise of any \emph{thermodynamic} glass transition \cite{berthier2011}. By contrast, gelation is related to metastable liquid-gas phase separation. Starting from this premise, we show that gelation is discontinuous while the approach to vitrification is continuous as a function of (effective) temperature or density. Our main result is that we can now provide clear physical measures that enable absolute discrimination between gels and glasses unlike the dynamical quantities considered previously \cite{zaccarelli2009,pham2006,koumakis2011}. This distinction is clearly shown in the relaxation time, local structural measures and (osmotic) pressure. The former two show a quasi-discontinuous jump upon gelation, and a continuous change upon the approach to vitrification, while the latter turns negative at gelation but rises approaching vitrification for hard spheres. Furthermore supercooled liquids and gels exhibit distinct local structure.

\section{Methods}
\label{sectionMethods}

We carry out confocal microscopy experiments where we track the colloids at the single-particle level. We use fluorescently labelled density and refractive index matched colloids. Despite screening, the electrostatic interactions can lead to an increase in \emph{effective} colloid volume fraction $\phi_{\rm e}$ compared to the absolute $\phi$ \cite{royall2013myth}. Here we use $\phi_{\rm e}$ throughout. 
We make the assumption that the full colloid-polymer system can be treated as an effective one-component colloid system. For the size ratios we consider, this is expected to be highly accurate \cite{dijkstra2000}. Thus the polymers are treated at the level of an attraction between the colloids. Since the polymers are integrated out, their contribution to the osmotic pressure does not directly appear. 
The experiments are mapped to event driven molecular dynamics (MD) simulations of a system with a square well interaction of width $0.03\sigma$ where $\sigma$ is the particle diameter and well depth $\epsilon$. The latter is set to be equivalent to polymer mass fraction $c_p$ in the experimental system, so we express state points in volume fraction $\phi_{\rm e}$ and $\epsilon$ or $c_p$. Despite differences in details of the interactions of the experiments and simulations, the observed behaviour is generic to systems with short-range attraction \cite{foffi2005}. 

\subsection{Experimental} 
\label{sectionExperimental} 

We used two systems, both of sterically stabilised polymethyl methacrylate colloids. The first system used particles of diameter $\sigma=2.40$ $\mu$m with polydispersity 4\%, as determined from static light scattering.  The polystyrene polymer used had a molecular weight $M_{w}=3.1\times10^{7}$, leading to a polymer-colloid size ratio of $q=0.18$ in good solvent conditions \cite{royall2007}. We used the first system for the path in the state diagram in Fig. \ref{figPhase} marked $(\overline{ba})$. Otherwise, we used a system with $\sigma=3.23$ $\mu$m, as determined from the first peak of $g(r)$ from confocal microscopy data and by scanning electron microscopy with 6\% polydispersity determined from SEM. The polymer had an $M_{w}=8.06\times10^{6}$ ($q=0.079$). 

In both cases the colloids and polymers were dispersed in a density- and refractive index-matching mixture of cis-decalin and cyclohexyl bromide. 4 mMol of tetrabutyl ammonium bromide salt was added to screen electrostatic interactions. Brownian times to diffuse a radius for the two systems were $\tau_B=3.11$ s and $7.59$ s. We find gelation at $c_{p}^{\rm gel}=1.0\pm0.2\times10^{-3}$ and $c_{p}^{\rm gel}=1.29\pm0.08\times10^{-4}$ for the ``gel'' and ``dense'' paths in Fig. \ref{figPhase} respectively. Samples were left for at least 30 minutes to ``equilibrate'', after which no change in structural or dynamic properties was seen on the experimental timescale (up to three days). 

The samples were imaged with a Leica SP5 confocal microscope fitted with a resonant scanner. Prior to imaging the samples were loaded into borosilicate glass capillaries (obtained from Vitrocom inc.) and sealed with epoxy resin. We allowed the resin 15 minutes to set prior to imaging and took data after letting the samples rest for a further 15 minutes. Imaging was carried out with the temperature fixed at 27 $^\circ$C with a temperature controlled stage and objective lens. During imaging, we saw no ageing of the samples. In other work, we have identified that initial remixing prior to arrest in gelation occurs on a timescale of less than one minute \cite{taylor2012}.

\subsection{Computer simulation.}
Event-driven molecular dynamics (MD) simulations were carried out with the DynamO package \cite{bannerman2011}. We used an equimolar five component mixture whose polydispersity is 8\%. Recent simulations suggest that larger polydispersity can influence the dynamics, but for the values we employ in our simulations and experiments are sufficiently monodisperse that all colloids arrest simultaneously \cite{zaccarelli2013}. For equilibrium simulations, we equilibrated for at least $10$ $\tau_{\alpha}$ and (unless otherwise indicated) sampled for at least a further $10$ $\tau_{\alpha}$ where $\tau_{\alpha}$ is the structural relaxation time. Simulation time was scaled to experimental data such that $\tau_{\alpha}$ for $\phi_{\rm e}\approx0.38$ was matched between both, namely, that $\tau_{\alpha}=2.597$ $\tau_{\rm B}$ which is then equivalent to the MD value of $0.404$ simulation time units. This scaling was applied throughout. Unless otherwise stated, for non-equilibrium state points we ran the system for at $10^5$ simulation time units prior to sampling. For our system sizes of 1372 particles and typical state points ($\phi=0.54$), this corresponds to a time of order $10$ minutes for the experimental system.

\subsection{Dynamics.}
We estimate the structural relaxation time $\tau_{\alpha}$ from the intermediate scattering function (ISF), $F(\mathbf{k},t)=\langle \sum_{j=1}^N \exp[i \mathbf{k} \cdot (\mathbf{r}(t+t')-\mathbf{r}(t'))] \rangle$ where the sum runs over all particles in the system. This we determine from coordinate data in the case of both experiments and simulations. The lengthscale upon which mobility is probed is set by the wavevector $k$ which here is taken to correspond to a particle diameter ($k \sim 2 \pi \sigma^{-1}$). The long-time tail of the ISF is fitted with a stretched exponential whose time constant is $\tau_{\alpha}$. The wavevector is taken close to the main peak in the static structure factor. The ISFs are shown as solid lines in Fig. \ref{figISF}\textbf{a}-\textbf{c} and light lines in \textbf{d}-\textbf{h}. For very deep quenches the ISF does not fully relax on the experimental or simulation timescale. In the case of simulation we run the system for $10^5$ simulation time units and sample for a further $10^5$ time units except in the ``very dense'' case (Fig. \ref{figISF}\textbf{h}). There we run the system for $5\time10^5$ time units and sample for a further $5\times10^5$ time units.

\subsection{Mapping state points between experiment and simulation} 
\begin{figure*}
\includegraphics[width=160mm]{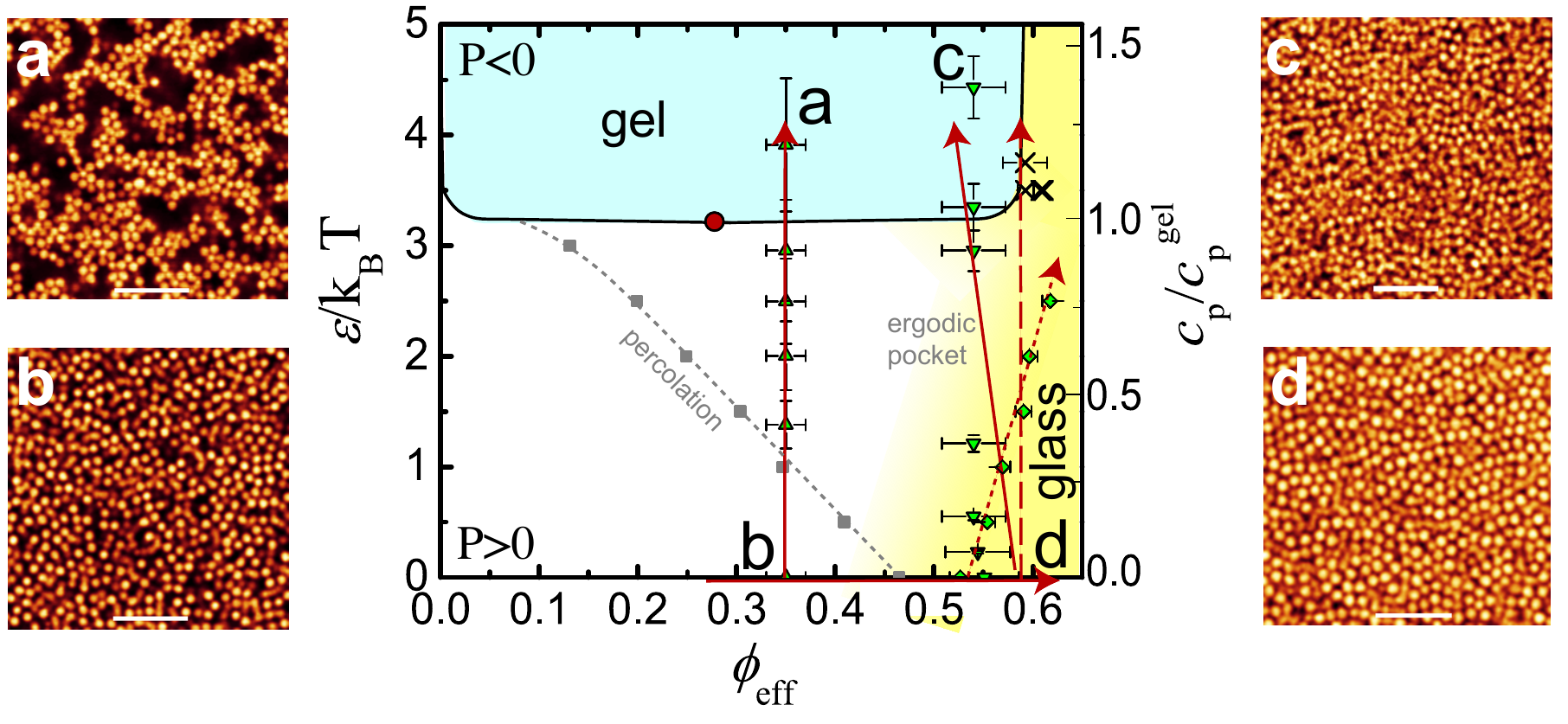}
\caption{{\bf State diagram of sticky spheres (3\% square well), with mapped experimental data.} $\varepsilon$ is well depth (inverse effective temperature), $c_{p}/c_{p}^{\rm gel}$ is the polymer concentration relative to that required for gelation and $\phi_{\rm e}$ is (effective) colloid volume fraction. Red circle is the critical point \cite{largo2008}. $p$ denotes the sign of (osmotic) pressure. Crosses indicate the volume fraction of the dense phase estimated from fully phase separated simulations. These are used to construct the liquid-gas phase separation line which delimits the gel region. Percolation is indicated as grey squares with a dashed line to guide the eye. Shaded areas indicate the onset of slow dynamics. Arrows denote different paths considered, with green symbols denoting experimental data points. $(\overline{bd})$ corresponds to the hard sphere glass transition, $(\overline{ba})$ (up triangles) to gelation at moderate $\phi_{\rm e}$, and $(\overline{dc})$ (down triangles) to gelation at high $\phi_{\rm e}$. The long dashed arrow refers to gelation at very high $\phi_{\rm e}=0.59$. The isobaric quench (diamonds) is shown as a dotted arrow. Confocal images of state points indicated in the phase diagram are as follows: \textbf{a,} ``normal gel'' ($\phi_{\rm e}=0.35$, $c_{p}/c_{p}^{\rm gel}=1.14$); \textbf{b}, ``hard spheres'' ($\phi_{\rm e}=0.35$); \textbf{c,} high-density gel ($\phi_{\rm e}=0.54$,$c_{p}/c_{p}^{\rm gel}=1.43$); \textbf{d,} ``hard spheres'' ($\phi_{\rm e}=0.58$). Here $c_{p}$ is the polymer mass fraction, and $c_{p}^{\rm gel}$ is the polymer mass fraction required for gelation. Bars=20 $\mu$m.
}
\label{figPhase} 
\end{figure*}

\begin{figure*}
\includegraphics[width=160 mm]{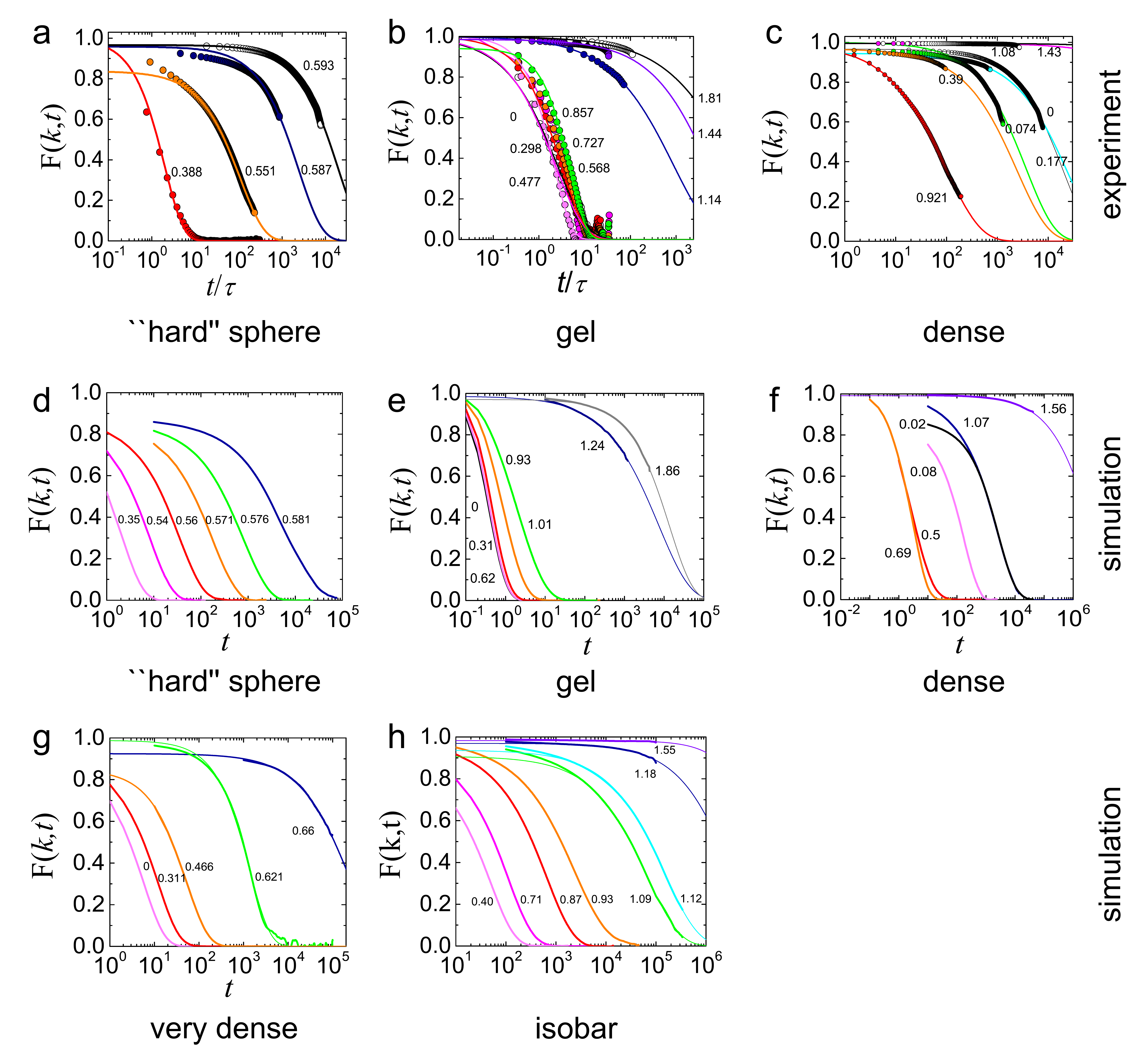}
\caption{Intermediate scattering functions. 
\textbf{a}-\textbf{c}, Intermediate scattering functions determined from experiments for the ``hard'' sphere $(\bar{bd})$ (\textbf{a}), gel $(\bar{ba})$ (\textbf{b}), and dense gel $(\bar{dc})$ (\textbf{c}) paths in Fig. \ref{figPhase}. \textbf{d}-\textbf{h,} MD data, for the ``hard'' sphere (\textbf{d}), gel (\textbf{e}), dense gel (\textbf{f}), very dense (\textbf{g}) and isobaric (\textbf{h}) paths. Experimental data are in units of Brownian time $\tau_{B}$, MD data are in simulation time units. In \textbf{a} and \textbf{d}, ISFs are shown at different $\phi_{\rm e}$, whereas in \textbf{b},\textbf{c} and \textbf{e}-\textbf{h} labels denote $c_{p}/c_{p}^{\rm gel}$. All ISFs are fitted with stretched exponentials to obtain $\tau_{\alpha}$. 
\label{figISF}
}
\end{figure*}

Despite screening the residual electrostatic charge with tetrabutyl
ammonium bromide salt, some repulsions between the colloids can remain. Although in the system with smaller particles [gelation $(\overline{ba})$ in Fig. \ref{figPhase}], the electrostatics are weak and neglected \cite{royall2007}, in the system with larger colloids which we use for the ``hard'' sphere $(\overline{bd})$ and dense $(\overline{dc})$ paths in Fig. \ref{figPhase}, we  
treat the electrostatic repulsions as a Yukawa interaction $u_Y(r)=\epsilon_{Y}\exp[-\kappa\sigma(r / \sigma -1)]/(r/\sigma)$. We set $\epsilon_{Y}=k_{B}T$ and the Debye length $\kappa^{-1}$ is taken as 100 nm, following \cite{royall2007}. We map the system to hard spheres, using the Barker-Henderson effective hard sphere diameter $\sigma_{e}=\intop_{0}^{\infty}dr[1-\exp(-\beta u(r))]$, where $\beta=1/k_{\rm B}T$ such that the effective packing fraction $\phi_{\rm e}=\phi(\sigma_{e}/\sigma)^{3}$. This increases $\phi_{\rm e}$ by around 8\% relative to the absolute packing fraction $\phi$ (note that in the event-driven molecular dynamics (MD) simulations, $\phi_{\rm e}=\phi$). 

Previously, we have shown that the Asakura-Oosawa (AO) potential describes the polymer-induced attractions between the colloids rather well \cite{royall2007}. Furthermore we have obtained good agreement with experiment by assuming simple addition of AO and Yukawa interactions for similar conditions to those we employ here \cite{royall2005}. We therefore make the same assumption that the colloid-colloid interaction is described by the sum of the AO and Yukawa interactions (for details see \cite{royall2007,royall2005}),  and map these to the square well model with well width 3\% and depth $\epsilon$, using the extended law of corresponding states \cite{noro2000}. This requires that the reduced second virial coefficient $B_2^*=3/\sigma_{e}^{3}\intop_{0}^{\infty}dr \ r^{2}[1-\exp(-\beta u(r))]$, is matched between the assumed experimental potential and the square well.  


\section{Results}
\label{sectionResults}

First we show the state diagram of our system in Fig. \ref{figPhase}. Here we use the effective volume fraction of colloids $\phi_{\rm e}$ and the polymer concentration $c_{\rm p}$ as the control parameters. The former changes the density and the latter changes the strength of the interparticle attraction. Each state point is prepared using the same experimental protocol see section \ref{sectionExperimental}. The crucial point is that the characteristic time of the protocol ($\sim 30$ min) is much longer than the timescale of demixing and the resulting dynamical arrest, but still much faster than the timescale of ageing. Because of this feature each state point is rather well-defined and may be compared with simulations. By connection to thermodynamics, it is possible to draw a line which distinguishes gels and glasses, under the premise above. This is the liquid-gas spinodal. However in this system it is effectively indistinguishable from the binodal separating the thermodynamically stable and unstable region. Its location is determined from literature data \cite{largo2008} and our simulations. Demixing brings the volume fraction of the more colloid-rich phase to extremely high values ($\phi_{\rm e} \approx0.59$), o that the dynamics are very slow. This formation of a very dense colloid-rich phase is common to any initial composition because of the very flat nature of the phase separation line. The important observation here is that across a very wide range of colloid volume fractions, one expects to cross the phase separation line at almost the same effective temperature, $c_{\rm p}$ (criticality), which occurs at $\epsilon^{*}=3.22$ $k_{\rm B}T$ for the $0.03\sigma$ square well \cite{largo2008}. In the experiments, we denote the polymer concentration required for gelation as $c_p^{\rm gel}$ which we take as the mid-point of the highest fluid $c_p$ and lowest gel $c_p$ as previously  \cite{royall2008g}. While ``liquids'' of spheres at $\phi_e\approx0.59$ are in principle metastable to crystallisation in our system this is totally suppressed by polydispersity.

To elucidate the different dynamical arrest scenarios, we consider the following paths through the state diagram indicated in  Fig. \ref{figPhase}. $(\overline{bd})$: increasing packing along the ``hard sphere'' line (with no added polymer) leads to vitrification; $(\overline{ba})$: at a relatively low colloid volume fraction ($\phi_{\rm e}=0.35$), addition of polymer results in gelation at a polymer concentration $c_p^{\rm gel}$ \cite{royall2008g}; $(\overline{dc})$: addition of polymer at high colloid density leads to a transition between two states with slow dynamics with fluid in between reminiscent of that observed previously \cite{pham2002}. This glass (Fig. \ref{figPhase}\textbf{d}) and gel (Fig. \ref{figPhase}\textbf{c}) are found at comparable volume fraction. This marks our first main finding: if the liquid-gas coexistence region of the phase diagram corresponds to gelation, far from being a low-density network, gels can be found at rather high volume fractions. We shall see below that gels have a local structure distinct from glasses at the same volume fraction, although both have similar disordered structures with homogeneous density on a longer lengthscale.

\vspace{1cm}
\noindent

\begin{figure*}
\includegraphics[width=160mm]{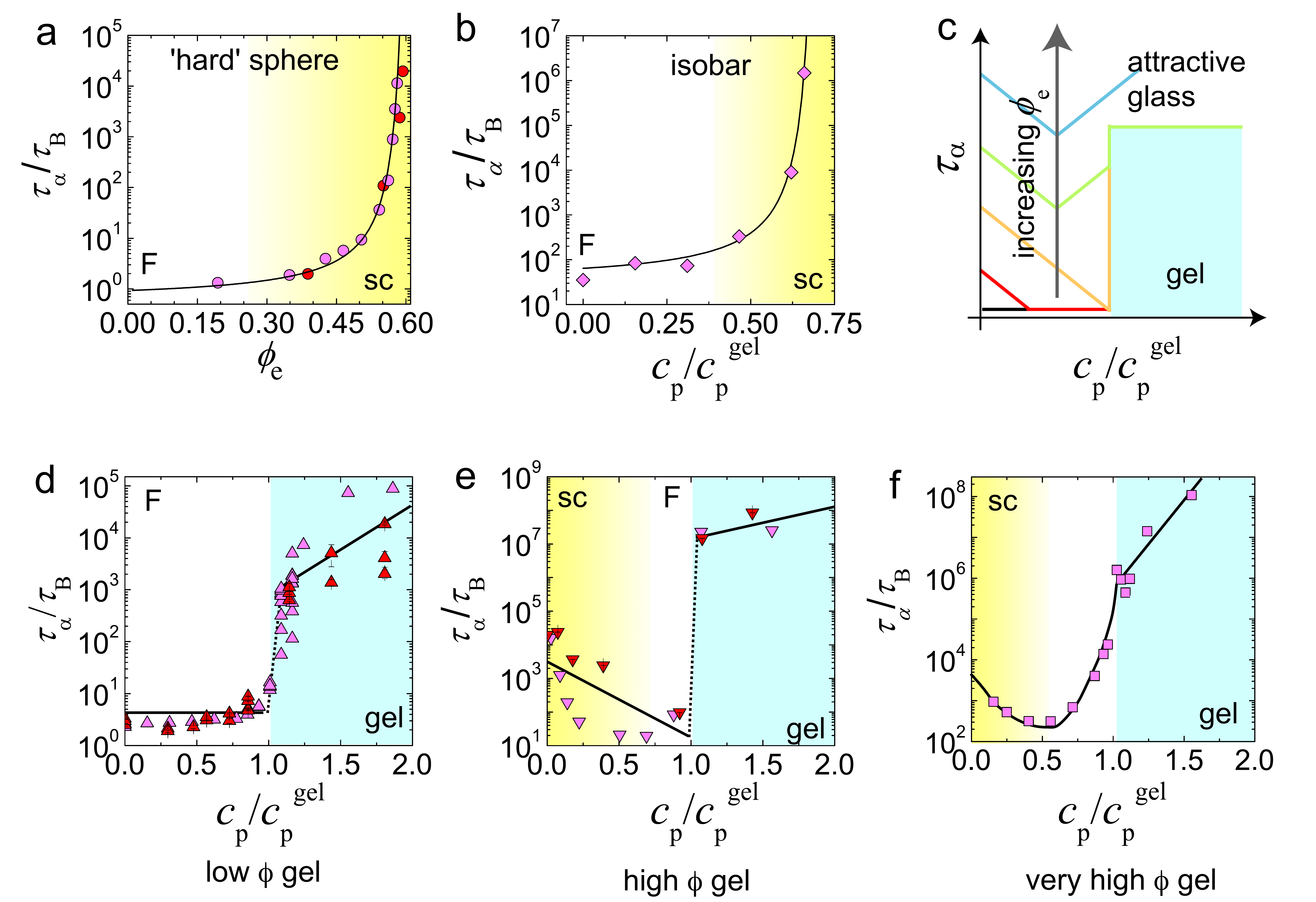}
\caption{{\bf Structural relaxation time $\tau_{\alpha}$ along the paths in Fig. \ref{figPhase} showing the continuous approach to vitrification and discontinuous gelation}. \textbf{a,} path $\overline{bd}$ indicating the continuous nature of the approach to the hard sphere glass. \textbf{b,} isobaric path showing no re-entrant dynamics. \textbf{c,} schematic of the gelation behaviour at different volume fractions is indicated by the coloured lines. \textbf{d-f,}  gelation at increasing volume fraction. \textbf{d:} $\phi_{\rm e}=0.35$, path $(\overline{ba})$; \textbf{e:} $\phi_{\rm e} \approx 0.54$, path $(\overline{dc})$; \textbf{f:} $\phi_{\rm e} = 0.59$, dashed path in Fig. \ref{figPhase}. In \textbf{a} and \textbf{b} the lines are VFT fits. In \textbf{c-f} the lines are to guide the eye. Experiment and simulation data are scaled such that ``hard'' spheres ($\phi_{\rm e}=0.38)$ agree. Red symbols are experimental data and pink are simulation. ``F'' and ``sc'' denote fluid and supercooled liquid respectively. $\tau_{\rm B}=\pi\eta\sigma^{3}/8k_{\rm B}T$, the time taken for a particle to diffuse its own radius at infinite dilution in a solvent of viscosity $\eta$. The error bars in the value for $\tau_\alpha$ are obtained from fitting the ISF data (see Sfig~S1). Here $\tau_B$ is the time to diffuse a radius.
}
\label{figDynamics} 
\end{figure*}

\subsection{Dynamics}

Next we focus on the dynamical behaviour associated with these paths. In Fig. \ref{figDynamics} we show the structural relaxation time  $\tau_{\alpha}$ (see Methods) along the paths in Fig. \ref{figPhase}. The hard sphere glass transition [path $(\overline{bd})$] exhibits a continuous increase in $\tau_{\alpha}$ as shown in Fig. \ref{figDynamics}a. Experimental and simulation data both exhibit very similar behaviour. The data are well described by the Vogel-Fulcher-Tammann (VFT) relation $\tau_{\alpha}=\tau_{0}\exp[D \phi_{\rm e}/(\phi_{0}-\phi_{\rm e})]$ where $\tau_{0}$ is a relaxation time in the normal liquid, $D$ is the ``fragility index'', and $\phi_{0}\approx0.61$ is the ideal glass transition volume fraction \cite{brambilla2009}. For $\phi_{\rm e}\gtrsim0.59$, structural relaxation does not occur on our timescales, so hard spheres at these high densities are glasses for our purposes.

This vitrification behaviour contrasts strongly with paths which cross the liquid-gas phase separation line (see Fig. \ref{figDynamics}d-f) \cite{angellComment}. All exhibit a discontinuity in $\tau_{\alpha}$ as the phase separation line is crossed at $c_p/c_p^{\rm gel}=1$. Gelation then is a discontinuous transition between a mobile fluid and an immobile gel which occurs at the phase separation line for any $\phi_{\rm e}$. This is consistent with previous work \cite{verhaegh1997,tanaka1999colloid,lu2008,pham2002} which identified gelation with (arrested) spinodal phase separation. Here we emphasise the dynamic manifestation of gelation --- a quasi-discontinuous jump in the relaxation time. This forms our second finding and constitutes the basis of the dynamic distinction between gelation and vitrification. Below we go further to explore the consequences of this spinodal gelation at high particle volume fraction.

The state diagram in Fig. \ref{figPhase} shows that gels can form at high volume fraction. Re-entrant dynamics along lines similar to path $(\overline{dc})$ have previously been identified with an attractive glass \cite{pham2002,poon2002,zaccarelli2002}. An ``attractive'' glass was predicted by mode-coupling theory (MCT) at high colloid volume fraction but the predicted attraction strength at arrest varies considerably from that found in simulation \cite{zaccarelli2002}, moreover distinguishing vitrification and gelation is not easy with MCT. 

Some experiments emphasise the role of percolation in arrest at moderate and high densities \cite{eberle2011} which is expected in the case that the bond lifetime is very long, but in this case the system remains ergodic even after percolation  \cite{tanaka1992}. Meanwhile others \cite{poon2002,grant1993,royall2008g} find no arrest at percolation which is the case here, as shown in Fig. \ref{figPhase}.  Here we see that for $\phi_{\rm e} \lesssim 0.59$, the phase separation line is crossed and hus a gel rather than an attractive glass is formed. This is evidenced by a discontinuous jump of three orders of magnitude in relaxation time upon crossing the phase separation line (see Fig. \ref{figDynamics}\textbf{e}). However for $\phi_{\rm e}$ greater than the dense side of the phase separation region, which we estimate as $\phi_{\rm e} \gtrsim 0.59$ we expect an attractive glass rather than a gel, as the line is not crossed. Thus it is possible that some work previously thought to pertain to a re-entrant glass transition in fact concerned gelation.
For $\phi_{\rm e} = 0.59$ in fact we find some dynamic slowing for $c_p/c_p^{\rm gel}<1$, i.e., in the one-phase region of the state diagram, as shown in Fig. \ref{figDynamics}f. In other words, at $\phi_{\rm e} = 0.59$, a supercooled liquid exhibits continuous slowing down characteristic of a glass with an increase in $\phi_{\rm e}$ before gelation intervenes. Further evidence that this path involves a thermodynamic transition is provided below with our measurements of (osmotic) pressure.

We summarise our findings of the relaxation time in the fluid-gel transition in Fig. \ref{figDynamics}c. At low densities, the system is mobile until  ($c_{p}/c_{p}^{\rm gel} \approx 1$), at which point it undergoes discontinuous dynamical arrest. Increasing colloid volume fraction leads to re-entrant dynamical behaviour as a function of increasing attraction once the volume fraction becomes sufficient that hard spheres exhibit slow dynamics. The fluid-gel transition remains abrupt. Upon increasing volume fraction further ($\phi_{\rm e}\rightarrow 0.6$), the gel transition gradually becomes masked by slow dynamics. We emphasize that both experiments and simulations show the same behaviour. Although equilibrium or metastable states might be reasonably well modelled by MD simulation \cite{berthier2007jpcm}, the nature of the dynamics can be important in gelation, particularly for low volume fractions \cite{furukawa2010}. Nevertheless, in Fig. \ref{figDynamics}\textbf{b} and \textbf{c}, we find good agreement between experiment and simulation. 

\begin{figure}
\includegraphics[width=85 mm]{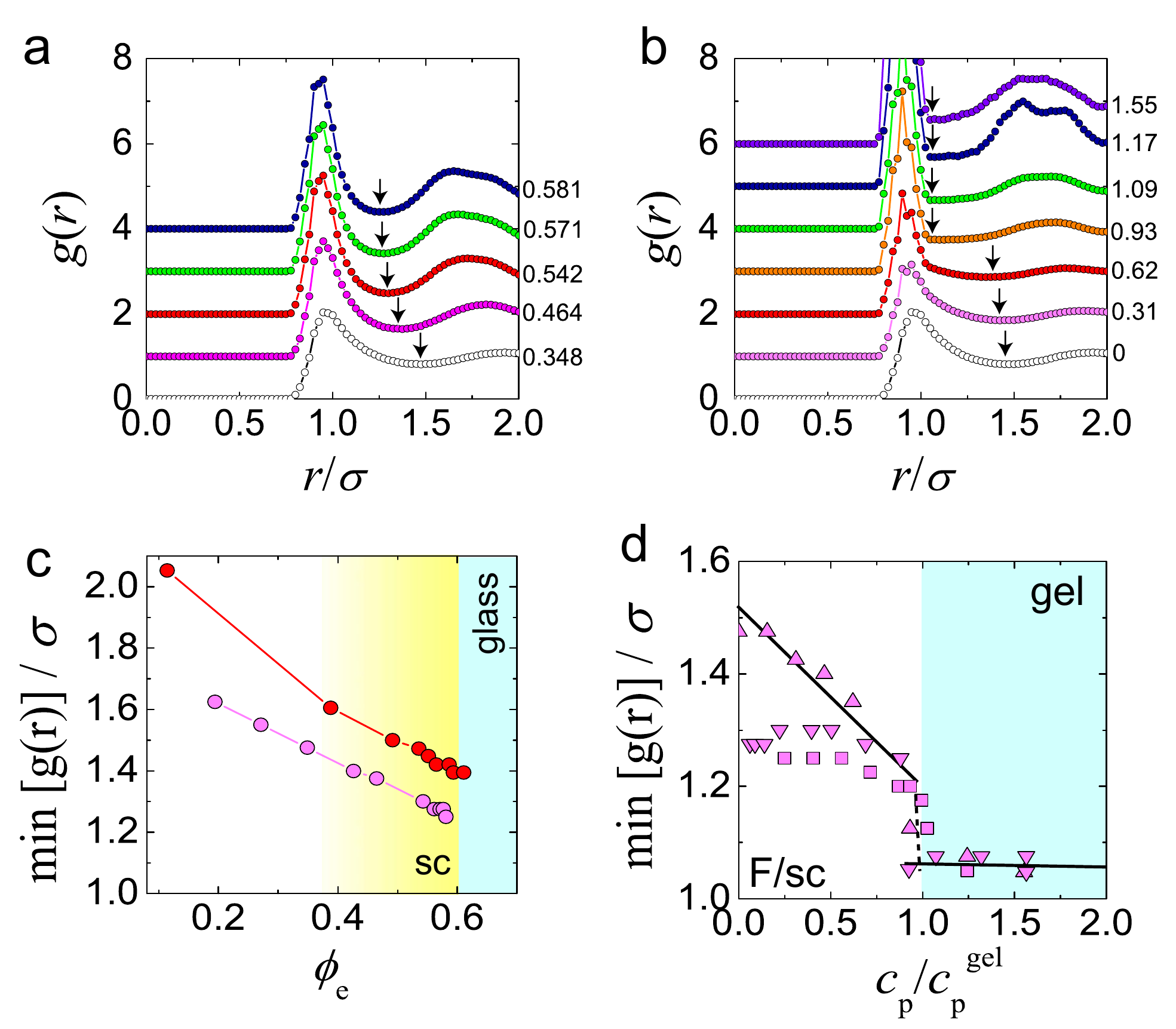}
\caption{\textbf{Distinguishing gelation and vitrification via the radial distribution function.} 
\textbf{a}, Radial distribution function $g(r)$ for different $\phi_{\rm e}$ for hard spheres with arrows indicating first minima. Data are labelled by $\phi_{\rm e}$. \textbf{b}, $g(r)$ along path $(\bar{ba})$ in Fig. \ref{figPhase} with arrows indicating the first minima. Data are labelled with $c_p/c_p^{\rm gel}$. \textbf{c,} the first minimum of $g(r)$ as a function of $\phi_{\rm e}$. \textbf{d,} the first minimum of $g(r)$ for state points along paths in Fig. \ref{figPhase}. The three paths indicated in Fig. \ref{figPhase} by  [$(\bar{ba})$, gel], [$(\bar{dc})$, dense gel], and (dashed line, very dense gel) are shown in \textbf{d} as down triangles, up triangles and squares, respectively. In \textbf{a,b}, data are offset for clarity. Red points in \textbf{c} are from experiments, otherwise all data are from simulations.
\label{figMoment}.}
\end{figure}

\begin{figure*}
\includegraphics[width=160mm]{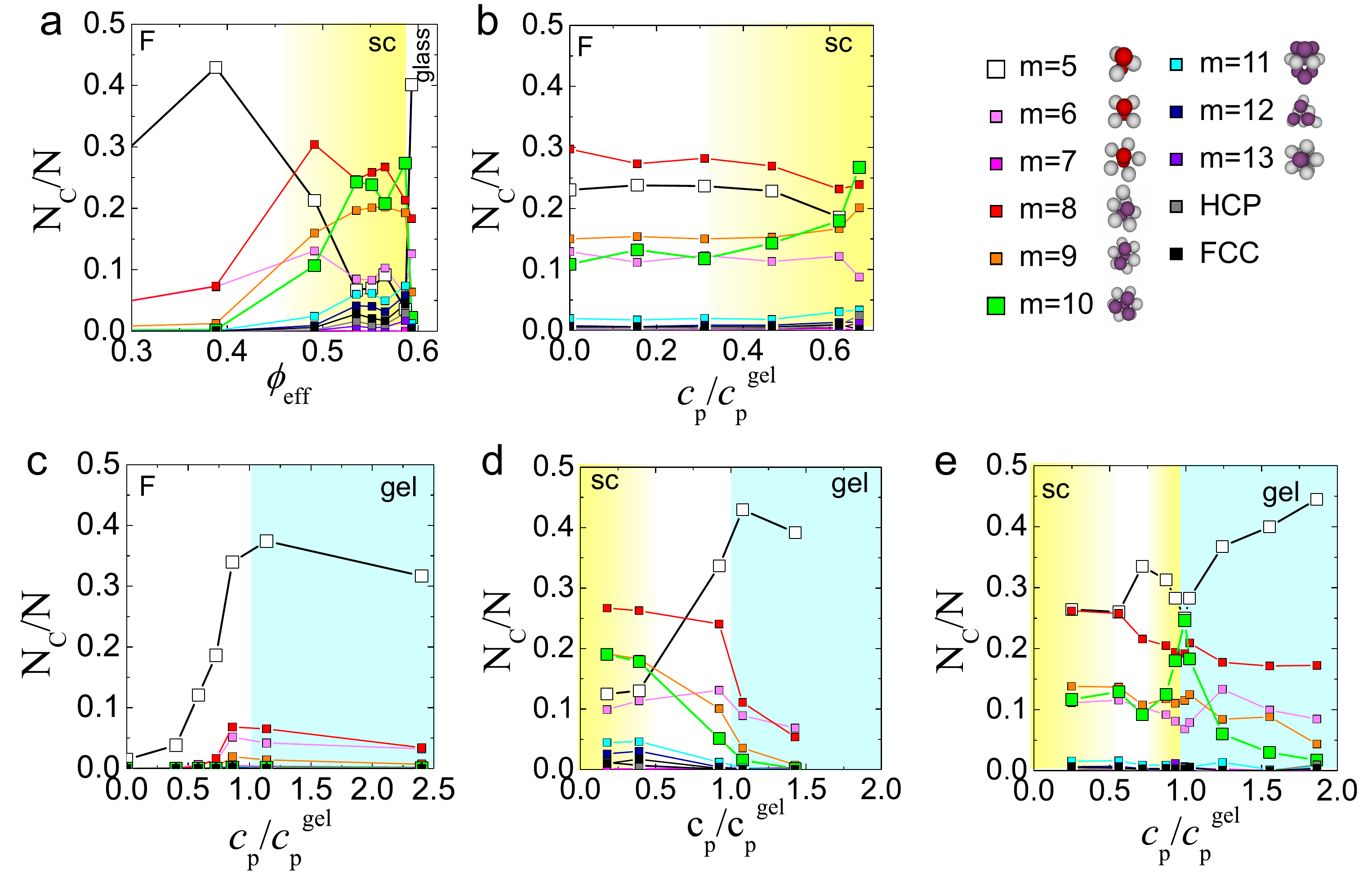}
\caption{{\bf Topological cluster classification of glass and gel transitions.} 
Cluster populations are plotted along paths in Fig. \ref{figPhase}. Vitrification along the paths in hard spheres $(\overline{bd})$ (\textbf{a}) and at constant pressure (\textbf{b}). These are contrasted with gelation \textbf{c-e}. \textbf{c}: $\phi_{\rm e} = 0.35$; \textbf{d}: $\phi_{\rm e} \approx 0.54$; \textbf{e:} $\phi_{\rm e} = 0.59$. Clusters considered in the TCC are indicated on right. Coloured lines correspond to different structures identified by the TCC and specified in the legend. sc denotes supercooled liquid.}
\label{figTCC} 
\end{figure*}

\begin{figure*}
\includegraphics[width=180mm]{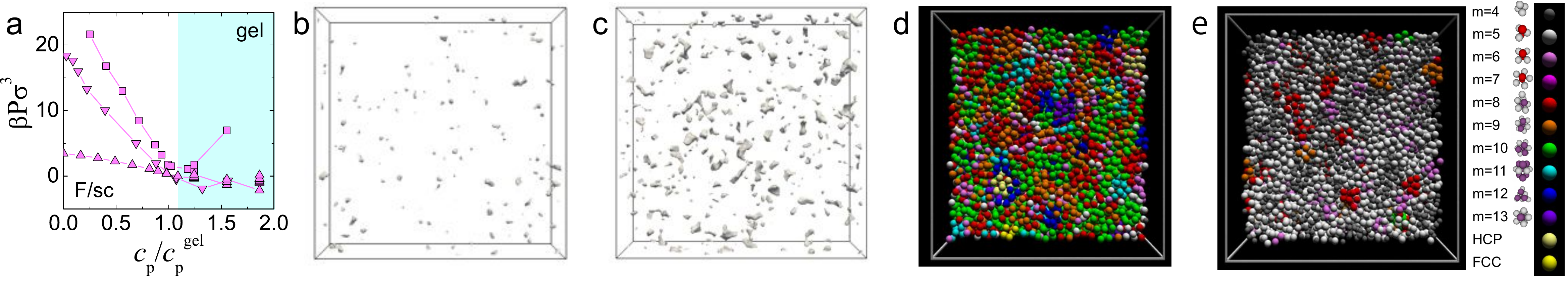}
\caption{{\bf  Pressure upon quenching and structural differences between glasses and gels at similar $\phi$. }
\textbf{a,} Pressure upon quenching sticky spheres at various packing fractions. Up triangles are ``gel'', path $(\bar{ba})$, down triangles ``dense gel'', path $(\bar{dc})$ and squares ``very dense'' (dashed line) Fig. \ref{figPhase}. MD simulation data. \textbf{b,} Voids in  ``hard'' spheres ($\phi_{\rm e}=0.54$).  \textbf{c,} Voids in dense gel ($\phi_{\rm e}=0.54$, $c_{p}/c_{p}^{\rm gel}=1.43$). \textbf{d,} and \textbf{e,} experimental data rendered following the topological cluster classification for  ``hard'' sphere supercooled liquid ($\phi_{\rm e}=0.59$) and dense gel ($\phi_{\rm e}\approx0.54$, $c_{p}/c_{p}^{\rm gel}=1.08$) respectively. Clusters considered in the TCC are indicated on right. }
\label{figStructure} 
\end{figure*}

\subsection{Isobaric quenches avoid gelation}

In the above, we have argued that gels should persist even up to colloid volume fractions $\phi_{\rm e} \gtrsim 0.59$, and now give further weight to this claim. If indeed a phase separation line is crossed, quantities such as the (osmotic) pressure may feature a discontinuity. Note that here we are considering an effective one-component system of colloids. Contributions from the polymer are integrated out and thus are not explicitly considered \cite{dijkstra2000}. From our simulations, we compute the pressure for the paths in Fig. \ref{figPhase}. As expected, discontinuities are found upon crossing the phase separation line, up to $\phi_{\rm e} = 0.59$ (pressure is shown in Fig. \ref{figStructure}\textbf{a}). In fact, for $\phi_{\rm e} \lesssim 0.54$, the pressure even turns negative. This provides evidence that the dense side of the phase-separation line lies at $\phi_{\rm e} \gtrsim 0.59$. The consequences for the re-entrant glass transition are considerable: for $\phi_{\rm e}\lesssim0.59$, gelation should intervene, which is supported by a discontinuous jump of $\tau_\alpha$ upon gelation (see Figs. \ref{figDynamics}\textbf{e} and \textbf{f}). In other words, for $\phi_{\rm e}\lesssim0.59$, the re-entrant glassy state becomes a gel when $c_{p}\gtrsim c_{p}^{\rm gel}$. Furthermore, $\phi_{\rm e}=0.59$  is a \emph{lower bound} for the high-density limit of gelation: in equilibrium, the ``liquid'' may be denser still.

That the (osmotic) pressure exhibits such a discontinuity leads to the hypothesis that constant-pressure paths through a phase diagram cannot produce a gel. We therefore carried out simulations at fixed pressure ($15.45$ $k_{\rm B}T\sigma^{-3}$) which led to the isobar indicated as a dotted line in Fig. \ref{figPhase}. Such a path should not produce a gel, that is, there should be no discontinuity, and in Fig. \ref{figDynamics}\textbf{b}, we indeed find a continuous increase in relaxation time as a function of attraction strength. The VFT fit in Fig. \ref{figDynamics}\textbf{b} indicates divergence at $c_{p}/c_{p}^{\rm gel}=0.709$, in contrast to the quenches at constant volume, all of which  showed a sudden slowdown in dynamics at $c_{p}/c_{p}^{\rm gel}\approx1$. Moreover, the isobar in Fig. \ref{figPhase} shows a departure to very high densities, so the system avoids gelation and undergoes vitrification. Put another way, at gelation (where the pressure tends to negative values) the system cannot support external stress (pressure). Now gels are found exclusively in soft materials formed of a mixture whose components have large size disparity  \cite{tanaka2000viscoelastic}.  Although in principle possible \cite{royall2011c60,testard2011}, molecular systems have not so far undergone spinodal gelation. Soft matter experiments are typically carried out at constant volume, while in experiments on molecular systems pressure is often fixed. The equivalent to isobaric quenching in our case would be constant osmotic pressure of the effective one-component colloid system. Were such experiments to be performed 
we expect gelation would not be observed. Our third finding is that gelation and (osmotic) pressure are intimately coupled. Fixing the pressure prevents the system from gelling, leaving vitrification as the only route of dynamical arrest available. This is directly linked to the fact that a gel is formed as a consequence of demixing, which is controlled by the {\it conserved} order parameter, i.e., the total volume fraction in the whole system, and the process takes place while conserving the composition $\phi_{\rm e}$.

\subsection{Structure}

The dynamics and phase behaviour provide a means by which gelation and vitrification may be distinguished. We shall now show that local structural measures also support the idea of gelation as a discontinuous transition to a state far from equilibrium. While it is well known that density-density correlations show little change upon vitrification \cite{ramos2005sdg}, here we find that gelation exhibits rather different behaviour. Crossing the phase separation line along the paths indicated in Fig. \ref{figPhase} leads to a discontinuity in the first minimum of the pair correlation function $g(r)$ which is entirely absent from the glass transition. 
Since gels are associated with demixing, upon crossing the phase separation line, we expect the first minimum of the pair correlation function $g(r)$ to move to a value close to contact ($\sigma$). Precisely this behaviour is found: in Fig. \ref{figMoment}\textbf{a} and \textbf{b} the first minimum of $g(r)$ is indicated by arrows for the hard sphere glass transition and gelation respectively. The latter case shows a clear jump around the phase separation line $c_p/c_p^{\rm gel}=1$, while for hard spheres, the minimum should and does decrease \emph{continuously} upon increasing density. The resulting $g(r)$ minima are then plotted as  as a function of $\phi_{\rm e}$ and $c_p/c_p^{\rm gel}$ in Fig. \textbf{c} and \textbf{d} respectively where they are labelled as $\min[g(r)]$. Discontinuous behaviour is seen at all $\phi_{\rm e}$ for the gels while in hard spheres, both experiments and simulations show a continuous fall in $\min[g(r)]$ as a function of $\phi_{\rm e}$.

The particle-level detail in our experiments also enables us to detect small voids inaccessible to other techniques. To make such voids visible, we apply a Gaussian blur of standard deviation $0.2\sigma$ to an image of spheres of diameter $\sigma$ reconstructed from coordinate data. As shown in Fig. \ref{figStructure}\textbf{c}, these are present in the gel, even though its high volume fraction ($\phi_{\rm e}=0.54$) prevents larger scale phase separation, but are almost absent from hard spheres at the same density (Fig. \ref{figStructure}\textbf{b}). Similar behaviour may be observed in Fig. \ref{figPhase}\textbf{c} and \textbf{d}.


\begin{figure*}
\includegraphics[width=160mm]{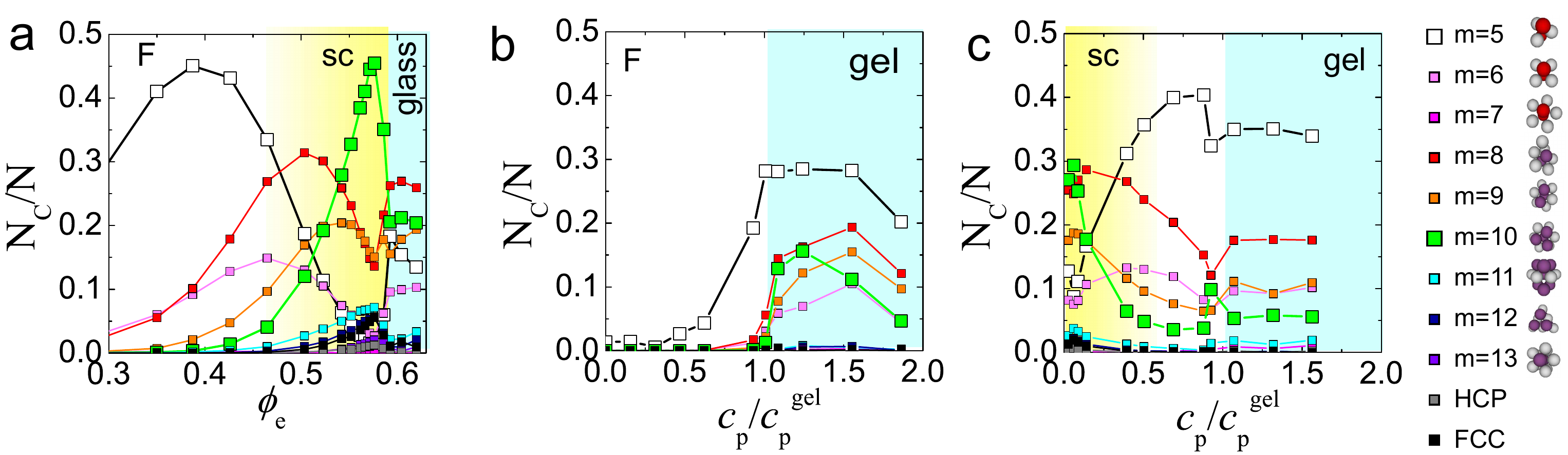}
\caption{{\bf Topological cluster classification  analysis for simulation data.} 
These data are mapped to the following paths in Fig. \ref{figPhase}, \textbf{a}, hard spheres; \textbf{b}, ``gel''; \textbf{c}, ``dense gel'', and correspond to the plots in Fig. \ref{figTCC}.}
\label{figTCCsim} 
\end{figure*}

To further probe the local structure we use the topological cluster classification (TCC), which identifies structures whose bond network is equivalent to clusters that minimise the potential energy in isolation \cite{malins2013tcc}. The TCC considers clusters of size $5\leq m\leq13$, along with HCP and FCC crystalline configurations of 13 particles. In the case of ``hard'' spheres, the relevant parameter is volume and the same set of clusters is appropriate to minimise the free volume \cite{manoharan2003,taffs2013}. For all systems we consider here, we use the clusters shown in Fig. \ref{figTCC}. We assign each particle to the largest cluster in which it is found \cite{royall2008g}.

For the TCC analysis, we use a simple bond criterion to define the bond network \cite{malins2013tcc,royall2008g}. Setting a bond length in this way is appropriate for dilute gels where particles in a network may have no close neighbours on one side (Fig. \ref{figPhase}\textbf{b}) \cite{royall2008g}. An appropriate bond length is the first minimum of $g(r)$. However our polydisperse systems necessitate a further constraint, namely that two large particles touch. This is readily implemented in simulation where the particle size is known. We use the same criterion for the experiments, noting that there the polydispersity is 0.25\% less and that therefore, although the particle size distribution is continuous, the number of large-large bonds lost is very small. Moderate changes in the bond length criterion, or indeed using a modified Voronoi construction with a maximum bond length \cite{malins2013tcc} had no impact upon our results. 

In the experimental coordinate data rendered in Fig. \ref{figStructure}\textbf{b}-\textbf{e}, particles are coloured according to the cluster in which they are identified by the TCC. It is immediately clear that the two states, supercooled liquid and gel, have a significantly different local structure, despite very similar $\phi_{\rm e}$. In Fig. \ref{figStructure}\textbf{d}, a considerable number of green particles associated with ten-membered clusters (point group symmetry $C_{3v}$) based on five-membered rings are found in the case of the supercooled liquid, while the gel (Fig. \ref{figStructure}e) is predominantly five-membered triangular biprisms formed from two tetrahedra (white). Thus supercooled liquids and gels have distinct local structure, even though they are at the same volume fraction and their overall structures apparently look similar. If we consider the response of the local structure to vitrification (Fig. \ref{figTCC}\textbf{a}), we see a gradual, continuous change up to the hard sphere glass transition ($\phi_{\rm e} \approx 0.59$). The $m=10$ cluster dominates the system just prior to vitrification. Simulations ( Fig.~\ref{figTCCsim}) show very similar behaviour. Interestingly, quenching along the isobaric path also shows an increase in the same $m=10$ cluster (Fig. \ref{figTCC}\textbf{b}).

By contrast, in the case of gelation at moderate density [path $(\overline{ba})$], Fig. \ref{figTCC}\textbf{c} shows a precipitous rise in cluster population. Not only is the rise in cluster population very sudden, but also the cluster involved is different: gels are dominated by the five-membered triangular biprism. The same holds for higher density in the case of path $(\overline{dc})$, although there the fluid is sufficiently dense that its own cluster population is considerable. Data from simulations of the ``gel'' and ``dense'' paths show similar behaviour (Fig.~\ref{figTCCsim}). Thus upon gelation, the larger clusters associated with denser fluids give way to $m=5$ clusters as discussed above. Remarkably, at $\phi_{\rm e}=0.59$, Fig. \ref{figTCC}\textbf{e} shows behaviour indicative of both vitrification and gelation. Just prior to gelation, there is a strong rise in the $m=10$ cluster associated with vitrification. However, for attractions greater than those required for gelation, the structure reverts immediately to the $m=5$ triangular bipyramid. This mirrors 
Fig. \ref{figDynamics}\textbf{f} where there there is some indication of glassy behaviour prior to gelation.

\subsection{Local structure far from equilibrium} 

\begin{figure*}
\includegraphics[width=120 mm]{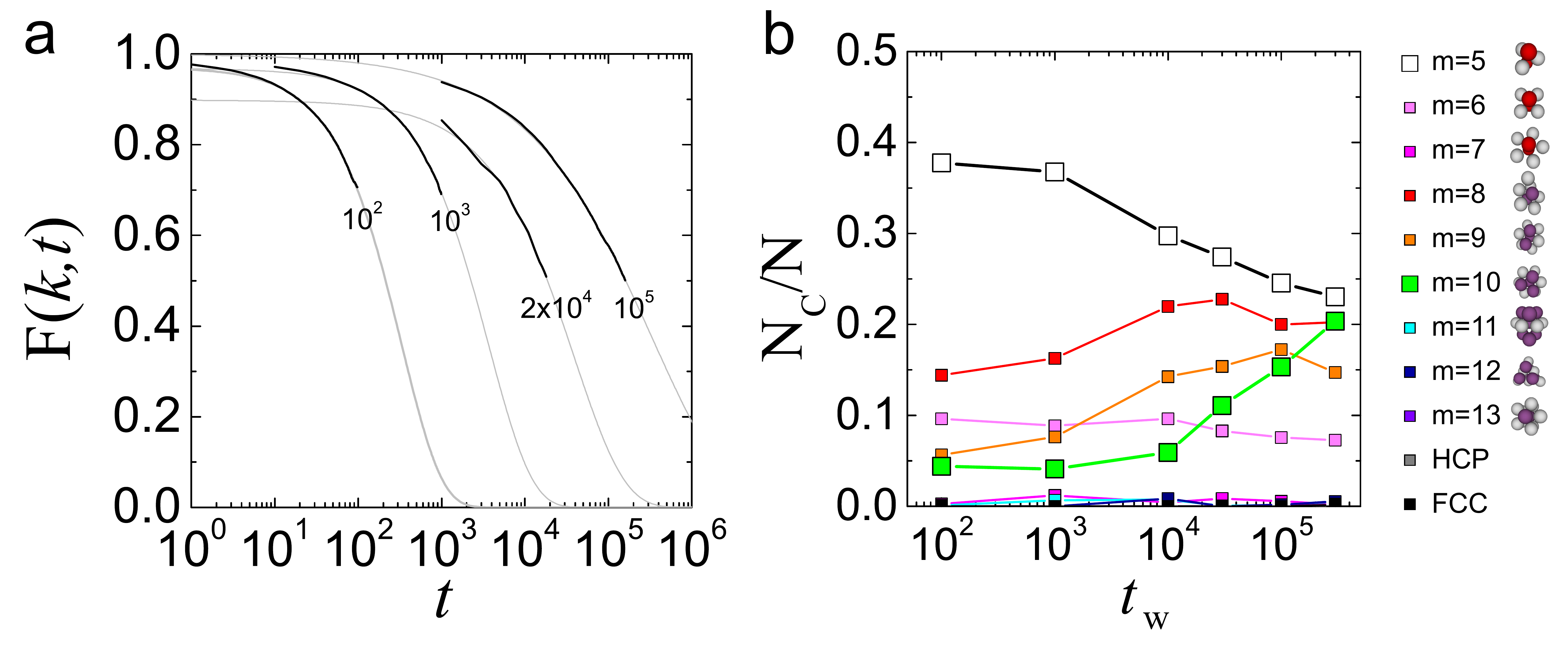}
\caption{\textbf{Ageing in gels.} \textbf{a} ISFs from MD data for $\phi_{\rm e}=0.35$ and $c_{p}/c_{p}^{\rm gel}=1.864$. Different waiting times are indicated by the number of collisions. Here the results from simulation (black lines) are fitted with a stretched exponential (grey lines) as described in the main text. In comparison with experimental data, we use a waiting time of at least $10^{5}$ simulation time units for sampling non-equilibrium data, which is comparable to the experimental waiting time. \textbf{b} TCC analysis of the structural evolution as a function of waiting time $t_w$.\label{figAgeing}}
\end{figure*}

What drives the system to select between $m=5$ and $m=10$ clusters? We offer the following explanation. At elevated packings, hard sphere like fluids exhibit high $m=10$ cluster populations \cite{taffs2013,taffs2010jcp}. By contrast due to rapid densification associated with phase separation, gels become arrested at times much shorter than the structural relaxation time. Thus the particles have no possibility to organise into larger $m=10$ clusters, but remain in the structure formed immediately upon compression. In particular, ageing can be sensitive to the dynamics : in equilibrium, our MD data might be expected to provide a reasonable description of the experiments, on timescales where particles have undergone sufficient collisions that their momentum has become uncorrelated. Out of equilibrium, the situation changes, and the nature of the dynamics can strongly influence the behaviour of the system \cite{furukawa2010}.  We therefore investigate the ageing behaviour of the system. Ageing is shown in Fig.~\ref{figAgeing}\textbf{a} for a gel in simulation with $\phi_e=0.35$ and $c_p/c_p^\mathrm{gel}=1.86$. We estimate that an equilibration time of 3000 simulation time units is comparable to the experimental waiting time. We see in the ISFs plotted in Fig.~\ref{figAgeing}a that longer waiting times indeed result in an increase in the structural decay time, but that this increase is around a factor of three for an order of magnitude increase in the waiting time. This lies within the scatter on Fig. \ref{figDynamics}\textbf{d} and \textbf{e} where experiment and simulation are compared. We saw no significant ageing effects in our experiments. In Fig.~\ref{figAgeing}\textbf{b} we show a TCC analysis for the structural evolution of the gel. Only at timescales greater than $3\times10^4$ time units is there a significant change in local structure. Such a change is not seen in experiments at comparable timescales (6 hours) but this may reflect differences in the dynamics. Thus at very long times, in simulation gels show a tendency towards $m=10$ clusters and so in gels some reorganisation can take place eventually.

This forms our fourth finding: local structure in highly non-equilibrium states is based on tetrahedra. Our reasoning is as follows. In a dense assembly of spheres, tetrahedra which are the basic arrangement of spheres in 3D, are formed at short times. 
This also holds for the $m=5$ clusters composed of two tetrahedra, whereas organisation of the tetrahedra into larger clusters is suppressed due to dynamical arrest. In other words the system is unable to reach a locally equilibrated configuration, for which formation of larger clusters is expected. These structural differences are clear in Fig. \ref{figTCC}, where the local structure of gels is dominated by triangular bipyramids while that of supercooled liquids shows many higher-order clusters, in particular $m=10$.

We can test this hypothesis that highly non-equilibrium states unable even to relax locally are dominated by $m=5$ triangular bipyramids by considering a hard sphere glass. If indeed, the inability to relax promotes these 5-membered triangular bipyramids, these should also dominate the glass --- unlike the supercooled fluid where relaxation is possible. This is precisely what we find, both in experiments (Fig. \ref{figTCC}a) and simulation (Fig. \ref{figTCCsim}\textbf{a}). This situation is similar to hyperquenching of molecular glassformers. Thus we identify a local structural motif to distinguish systems where local relaxation can occur and where it cannot. This can be used as a structural measure for how far from equilibrium a system is.

We now speculate on the consequences of our work for other systems. 
It is known that polymer solutions \cite{hikmet1988}, clay suspensions \cite{tanaka2004nonergodic}, and protein solutions \cite{cardinaux2007,gibaud2010} 
show spinodal gelation.  
All these systems have strong dynamic asymmetry between the two components, and accordingly all exhibit viscoelastic phase separation similarly to colloidal suspensions described here \cite{tanaka2000viscoelastic,tanaka1999colloid,tanaka2004nonergodic}, indicating that our findings may also apply to these systems. 
On the other hand, associating polymers \cite{tanaka1992} and emulsions with telechelic polymers \cite{filiali2001} and DNA {biffi2013} can also form gel by percolation without accompanying phase separation. These examples provide information on 
what kind of gel-forming systems, our scenario 
may apply. We argue that it should apply to any systems that exhibit viscoelastic phase separation, which produces the slow-component-rich phase beyond the glass-forming composition. 
This point needs to be checked carefully in the future. 

\section{Conclusion} 
\label{sectionConclusions}
 
In conclusion, we have shown that gelation and vitrification in an important model system may be distinguished by the quasi-discontinuous nature of the former and the continuous nature of the latter as a function of density and attraction, which controls the dynamics. The discontinuity in gelation manifests itself both in dynamical and structural properties, namely, the relaxation time $\tau_{\alpha}$ and structural measures using the topological cluster classification (TCC). Underlying this behaviour is that gelation involves a thermodynamic transition (phase separation), which leads to rapid densification and the resulting hyperquenching of a system.

Thus, gelation and vitrification are readily distinguished in sticky spheres by measuring dynamical or structural quantities along paths in the $(\phi_{\rm e},\varepsilon)$ plane. Moreover the TCC reveals a clear structural signature of systems which fail to relax. In the case of gels of sticky spheres and rapidly compressed hard sphere glasses, the structure is dominated by small tetrahedral and triangular bipyramid clusters, while supercooled liquids exhibit larger ten-membered clusters which form at times longer than the structural relaxation time. That the larger ten-membered structures require structural relaxation to form in turn means that the smaller structures indicate ``hyperquenching'' or states very far from equilibrium. Since this disorder is related to rapid densification upon (arrested) phase separation, colloidal gels should --- and do --- share local structural properties with rapidly compressed hard sphere glasses. Both are dominated by small tetrahedral and triangular bipyramid clusters. This poses an interesting fundamental question of whether the difference remains after long-time ageing. Although we could not access such states, we speculate that both tends to approach a similar state towards 
an equilibrated state with more fivefold symmetric structures.

Our work reveals a significant change in the local structure when a supercooled liquid falls out of equilibrium. Our basic geometric argument, that relaxation leads to an exploration of structures which locally minimise the free energy, rather than the tetrahedra formed upon compression, suggests that this may have some universality which could be explored in other glass-forming systems such as metallic glasses.

We have also found that the gel state persists up to unexpectedly high volume fractions, at least to $\phi\gtrsim0.59$, and may reside at higher density still. Therefore, gels need not take the form of networks. Our work suggests that the attractive glass may be found only at densities $\phi\gtrsim0.59$. 
It is an intriguing question how to distinguish repulsive and attractive glasses from their structures.
We have observed phase separation with the unusually high ``liquid'' volume fraction of 0.59. This prompts further work to explore just how dense the liquid state can become. 
Moreover, we have highlighted the role of pressure in gelation, and argued that quenching at constant volume is necessary for gelation and explains its prevalence in soft matter. This can be rephrased by that phase demixing characteristic of the conserved order parameter, i.e. the colloid volume fraction, is necessary for the colloidal gel formation.  This motivates experiments where osmotic pressure is fixed to test this prediction. Such experiments may be done by using a membrane through which solvent molecules pass. 



\begin{thebibliography}{43}
\bibitem[{ang()}]{angellComment}
\bibinfo{note}{Although these plots have similar axes to ``Angell plots'', we
  do not suggest an analogy to fragile or strong liquids. Rather this is a
  suitable representation to show the discontinuity in the dynamics.}
\end{thebibliography}

\vspace{2cm}
\noindent \textbf{Acknowledgements} 
The authors would like to thank Bob Evans and Rob Jack for helpful discussions, and Marcus Bannerman for help with the DynamO simulations, and to Mathieu Leocmach for help with experimental analysis and figure preparation. C.P.R. acknowledges the Royal Society, the European Research Council (ERC consolidator grant NANOPRS, project number 617266), the Japan Society of the Promotion of Science for financial support and EPSRC grant code EP/H022333/1 for the provision of a confocal microscope. H.T. acknowledges Grants-in-Aid for Scientific Research (S) and Specially Promoted Research from JSPS and also by Aihara Project, the FIRST program from JSPS, initiated by CSTP.





\end{document}